\documentclass[pdfa,a4paper,USenglish]{oasics-v2021}
\pdfoutput=1
\nolinenumbers

\usepackage{mathpartir}
\usepackage{nccmath}
\usepackage{relsize}
\usepackage{stmaryrd}
\usepackage{tikz}
\usetikzlibrary{arrows.meta}
\usetikzlibrary{backgrounds}
\usetikzlibrary{calc}
\usetikzlibrary{positioning}
\usetikzlibrary{quotes}
\usetikzlibrary{shapes.multipart}

\newcommand\rulename[1]{\textsc{\footnotesize #1}}

\newlength\mathgap
\newcommand\alt{\mathrel{\:\mid\:}}
\newcommand\defeq{\mathrel{\,\triangleq\,}}

\newcommand\edge[3]{#1 \rightarrow #2}

\newcommand\graphedges[1]{\mathcal{E}_{#1}}
\newcommand\graphnode[2]{\mathcal{N}_{#1}^{#2}}
\newcommand\nilscopestack{\mathit{\circ}}
\newcommand\nilsymbolstack{\mathit{\diamond}}
\newcommand\nodefile[2]{\mathcal{F}_{#1}^{#2}}

\newcommand\pathstep[1]{\textsf{(#1)}}

\newcommand\spathstep[1]{\textsf{\smaller (#1)}}
\newcommand\stackgraph[1]{#1}
\newcommand\stackgraphpath[4]{#1 \leadsto #2~\{#3\}}
\newcommand\stackgraphpartialpath[6]{\{#2\}\;#1 \leadsto #4\;\{#5\}}
\newcommand\sym[1]{\texttt{#1}}
\newcommand\validpath[2]{#1 \vdash #2}

\newcommand\sid[1]{\ensuremath{_\mathrm{#1}}}
\newcommand\OldSymbol[2]{\ensuremath{\texttt{#1}\sid{#2}}}
\newcommand\PseudoSymbol[1]{\ensuremath{\texttt{#1\strut}}}
\newcommand\InternalScope{\ensuremath{\mathit{\tikz[baseline=-3pt]{
    \node[int scope, semithick, minimum size=0.35cm] (char) {};}}}}
\newcommand\ExternalScope[1]{\ensuremath{\mathit{\tikz[baseline=-3pt]{
    \node[int scope, semithick, minimum size=0.35cm] (char) {#1};}}}}

\newcommand\RootNode{\ensuremath{\mathit{\tikz[baseline=-3pt]{\node[root node] (char) {};}}}}
\newcommand\Class[1]{\textsf{CLASS}(#1)}
\newcommand\Instance[1]{\textsf{INST}(#1)}
\newcommand\Num{\textsf{NUM}}
\newcommand\Rec[1]{\textsf{MOD}(#1)}
\newcommand\PushSymbol[1]{\ensuremath{\mathit{\textrm{\raisebox{0.05em}{\scriptsize $\downarrow$}} \; #1}}}

\newcommand\PopSymbol[1]{\ensuremath{\mathit{\textrm{\raisebox{0.05em}{\scriptsize $\uparrow$}} \; #1}}}

\newcommand\cyn{\nilsymbolstack}
\newcommand\cye{x}
\newcommand\cys{\hat{x}}

\newcommand\ccn{\nilscopestack}

\newcommand\ccs{\hat{i}}

\newcommand\pyn{\nilsymbolstack}

\newcommand\pyv{\psi}
\newcommand\pyp{\hat{x}}
\newcommand\pys{\Psi}

\newcommand\pcn{\nilscopestack}

\newcommand\pcs{\Phi}

\newcommand\lift[1]{\llbracket #1 \rrbracket}

\newcommand\git{\textsf{git}}
\newcommand\ts{\textsf{tree-sitter}}
\newcommand\tsg{\textsf{tree-sitter-graph}}

\title{Stack graphs}
\subtitle{Name resolution at scale}

\author{Douglas A. Creager}{GitHub, United States \and \url{https://dcreager.net}}{dcreager@dcreager.net}{https://orcid.org/0000-0003-1100-4894}{}
\author{Hendrik van Antwerpen}{GitHub, Netherlands \and \url{https://hendrik.van-antwerpen.net}}{hendrik@van-antwerpen.net}{https://orcid.org/0000-0001-5117-0921}{}

\authorrunning{D.\,A. Creager and H. van Antwerpen}

\Copyright{Douglas A. Creager and Hendrik van Antwerpen}

\keywords{Scope graphs, name binding, code navigation}

\ccsdesc[500]{Theory of computation~Program analysis}
\ccsdesc[500]{Software and its engineering~Software libraries and repositories}

\supplementdetails[cite=StackGraphs]{Software}{https://github.com/github/stack-graphs/}
\supplementdetails[cite=TreeSitterGraph]{Software}{https://github.com/tree-sitter/tree-sitter-graph/}

\newcommand\lstfilename[1]{\textsf{\footnotesize #1}}
\tikzset{node distance=0.4cm}

\tikzset{root node/.style={
    circle, fill=black, inner sep=0pt,
    minimum size=0.35cm,
}}

\tikzset{jump node/.style={
    circle, inner sep=0pt, minimum size=0.35cm,
    draw=black, thick, fill=white,
    path picture={
        \node at (path picture bounding box.center)
        [anchor=center, circle, fill=black, minimum size=0.2cm] {};
    },
}}

\tikzset{drop node/.style={
    circle, inner sep=0pt, minimum size=0.35cm,
    draw=black, thick, fill=white,
    path picture={
        \draw [black, thick]
        (path picture bounding box.north west) --
        (path picture bounding box.south east)
        (path picture bounding box.north east) --
        (path picture bounding box.south west);
    },
}}

\tikzset{int scope/.style={
    circle, fill=white,
    draw=black, thick, text=black, node font={\sffamily \smaller[2]},
    minimum size=0.45cm, inner sep=0em,
}}

\tikzset{old symbol/.style={
    fill=white,
    draw=black, thick,
    inner xsep=4pt,
    text height={0.7\baselineskip},
    text depth={0.3\baselineskip},
    node font={\relscale{0.8}},
}}

\tikzset{sg symbol/.style={
    fill=white,
    draw=black, thick,
    inner xsep=4pt,
    text height={0.7\baselineskip},
    text depth={0.3\baselineskip},
    node font={\relscale{0.8}},
    rectangle split,
    rectangle split parts=2,
    rectangle split horizontal,
    rectangle split draw splits=false,
    rectangle split ignore empty parts,
}}

\tikzset{pseudo symbol/.style={sg symbol, densely dashed}}

\definecolor{neronpath}{rgb}{0.40,0.40,0.41}

\tikzset{neron path/.style={
    draw=neronpath,
    text=neronpath,
}}

\tikzset{neron path resolved/.style={
    draw=neronpath,
    text=neronpath,
    densely dashed,
}}

\tikzset{neron/.tip={Triangle}}
\tikzset{neron import/.tip={Triangle[open,fill=white]}}
\tikzset{neron path/.tip={Triangle[fill=neronpath]}}
\tikzset{neron path import/.tip={Triangle[fill=neronpath, open]}}

\tikzset{hendrik/.tip={Triangle}}
\tikzset{hendrik link in/.tip={Square}}
\tikzset{hendrik link out/.tip={To}}
\tikzset{hendrik label/.style={
    fill=white, inner sep=0.2em,
    node font={\sffamily \smaller[2]},
}}
\tikzset{hendrik side label/.style={
    inner sep=0.3em,
    node font={\sffamily \smaller[2]},
}}

\tikzset{sg/.tip={Stealth[length=5pt,width=4pt]}}
\tikzset{sg path/.style={line width=1.8pt}}

\definecolor{codea}{rgb}{0.93,0.93,0.94}
\definecolor{codeb}{rgb}{0.80,0.80,0.81}
\definecolor{codeoutlinea}{rgb}{0.70,0.70,0.71}
\definecolor{codeoutlineb}{rgb}{0.60,0.60,0.61}

\tikzset{shade a/.style={draw=codeoutlinea, fill=codea, rounded corners}}
\tikzset{shade b/.style={draw=codeoutlineb, fill=codeb, rounded corners}}

\makeatletter
\AtBeginEnvironment{figure}{\def\@floatboxreset{%
    \setlength{\abovedisplayskip}{0pt}%
    \setlength{\abovedisplayshortskip}{0pt}%
    \setlength{\belowdisplayskip}{0pt}%
    \setlength{\belowdisplayshortskip}{0pt}%
}}
\makeatother

\EventEditors{Ralf L\"{a}mmel, Peter D. Mosses, and Friedrich Steimann}
\EventNoEds{3}
\EventLongTitle{Eelco Visser Commemorative Symposium (EVCS 2023)}
\EventShortTitle{EVCS 2023}
\EventAcronym{EVCS}
\EventYear{2023}
\EventDate{April 5, 2023}
\EventLocation{Delft, The Netherlands}
\EventLogo{}
\SeriesVolume{109}
\ArticleNo{8}
\category{Research}

\begin{document}

\maketitle

\begin{abstract}
    We present \emph{stack graphs}, an extension of Visser et al.'s scope graphs
    framework. Stack graphs power Precise Code Navigation at GitHub, allowing
    users to navigate name binding references both within and across
    repositories. Like scope graphs, stack graphs encode the name binding
    information about a program in a graph structure, in which paths represent
    valid name bindings. Resolving a reference to its definition is then
    implemented with a simple path-finding search.

    GitHub hosts millions of repositories, containing petabytes of total code,
    implemented in hundreds of different programming languages, and receiving
    thousands of pushes per minute. To support this scale, we ensure that the
    graph construction and path-finding judgments are \emph{file-incremental}:
    for each source file, we create an isolated subgraph without any knowledge
    of, or visibility into, any other file in the program. This lets us
    eliminate the storage and compute costs of reanalyzing file versions that we
    have already seen. Since most commits change a small fraction of the files
    in a repository, this greatly amortizes the operational costs of indexing
    large, frequently changed repositories over time. To handle type-directed
    name lookups (which require “pausing” the current lookup to resolve another
    name), our name resolution algorithm maintains a stack of the currently
    paused (but still pending) lookups. Stack graphs can be constructed via a
    purely syntactic analysis of the program's source code, using a new
    declarative \emph{graph construction language}. This means that we can
    extract name binding information for every repository without any
    per-package configuration, and without having to invoke an arbitrary,
    untrusted, package-specific build process.
\end{abstract}

\section{Introduction}

Code editors have long provided productivity features like \emph{code
navigation}, which let the user see and navigate the structural relationships in
their code—specifically which entities the names in their code refer to. These
features are equally useful on a software forge like
\href{https://github.com/}{GitHub}. However, large software forges must support
a completely different magnitude of scale, along a number of axes: the total
volume of stored code; the number of changes received over time; the need to
query historic versions of the code; the number of users making queries
simultaneously; and the number of programming languages to support.

We have developed \emph{stack graphs} to address these constraints. Stack graphs
build on Visser et al.'s \emph{scope graphs} framework \cite{Zwaan2023}, which
use a graphical notation to encode the name binding rules for a programming
language. Like scope graphs, each name binding in a program is represented by a
path in the corresponding graph. Unlike scope graphs, our path-finding algorithm
maintains a \emph{stack} of pending queries (hence the name), allowing a name
binding to depend on the results of other, intermediate name bindings.
Stack graphs are \emph{file-incremental}, meaning that each file in the source
program is represented by a disjoint subgraph, with no edges crossing between
them. (Special \emph{virtual edges} are created at query time to cross between
files.) We can construct each file's subgraph in isolation, without inspecting
any of the other files in the program. This lets us easily detect and reuse
results for file versions that we have already seen and analyzed. Since most
commits in a project's history change a small fraction of files in the project,
this greatly reduces the computational and storage costs of providing this
service.

This paper is organized as follows. In §\ref{sec:background}, we provide an
overview of scope graphs and the limitations that led to the development of
stack graphs. In §\ref{sec:stack-graphs} and §\ref{sec:partial-paths}, we
describe the stack graphs formalism in detail. In §\ref{sec:discussion}, we
discuss related work. In §\ref{sec:conclusion}, we conclude.

\section{Background and historical perspective}
\label{sec:background}

We first started investigating how best to add code navigation capabilities to
GitHub in 2017. Code navigation is widely available in local editors, so an
appealing potential approach would be to adopt whatever technology powers the
local editor solution, porting it as necessary to run in our production web
server environment.

However, there are clear differences between the code navigation experience in a
local editor compared with a software forge. Unlike the local editor, where
there is a single user viewing and interacting with a single project (and
primarily a single current version of the project's code), a software forge must
support an unbounded number of users simultaneously viewing any historic version
of any of the code hosted on the forge. Moreover, we must support users who are
\emph{exploring} and \emph{browsing} the code, in addition to those who are
\emph{authoring} and \emph{maintaining} it. These explorers often view code
months or years after it has been written. This means that we must either keep
our code navigation data available, and readily queryable, indefinitely; or
ensure that we can (quickly) regenerate that data on demand in response to a
future user query. This leads to a clear distinction between \emph{index time},
when a user “pushes” a new snapshot of a program or library to the forge, and
\emph{query time}, when a (likely different) user wants to view or explore that
snapshot at some point in the future.

We have ambitious but conflicting latency goals for the index and query phases.
Our most important goal is to minimize the delay between a user performing a
code navigation query and us displaying the results of that query in their
browser \cite{Miller1968}. To achieve this, we must spend some time
precalculating information at index time. That said, we do not want to perform
\emph{too much} precalculation work, so that code navigation is available
quickly after each push,\footnote{For querying, our goal is to show results in
under 100~ms. For indexing, our less precise goal is that code navigation should
be available before the user can \textsf{Alt-Tab} from their \git\ client over
to the browser.} and so we do not waste time precalculating code navigation data
for a repository or commit that will never be viewed.

Our scale also presents unique challenges. The most obvious is the volume of
code: GitHub holds petabytes of code history, and receives thousands of new code
snapshots each minute. It is imperative that our framework be
\emph{incremental}, skipping the costs of reanalyzing code that we have already
seen. Luckily, much of each project's history is redundant, with most commits
changing only a small fraction of files. As such, we particularly prefer
\emph{file-incremental} analyses, where we can analyze each source file at index
time in isolation, without inspecting (or having access to) any of the other
files in the project. The Merkle tree \cite{Merkle1987} data model used by \git\
provides a unique \emph{blob identifier} for each file version, which depends
only on the file's content. Because a file's identifier is available before
analysis starts, we can skip the storage \emph{and computation} costs of
redundant file-incremental work.

Lastly, GitHub hosts code written in a staggering number of programming
languages.\footnote{As of this writing, Linguist \cite{Linguist}, our
open-source language detection library, includes over 500 distinct languages and
sublanguages in its ruleset.} We want code navigation to work for all of them.
It will never be cost-effective for GitHub engineers to implement support for
the long tail of less popular languages. But there must be a path for
\emph{someone}---whether a GitHub engineer or a member of an external language
community---to add support for every programming language that exists and is in
use.

As we elaborate in §\ref{sec:discussion}, existing local editor code navigation
solutions do not satisfy these constraints. While researching whether existing
academic work could help tackle this problem, we discovered the \emph{scope
graphs} framework \cite{Zwaan2023}, which
introduced a novel and intuitive approach for encoding the name binding
semantics of a programming language in a graph structure.\footnote{Note that our
initial discovery and investigation of the scope graphs framework predates
van~Antwerpen joining GitHub.} This seemed likely to satisfy our language
support goals: all language-specific logic would be isolated to the graph
construction process, and all querying would happen via a single
language-independent algorithm that operated on that graph structure.
However, we quickly discovered that scope graphs, as originally described, would
not meet our scale requirements. In the rest of this section, we will explore
why, using a simple Python program as a running example.

\subsection{Néron scope graphs}

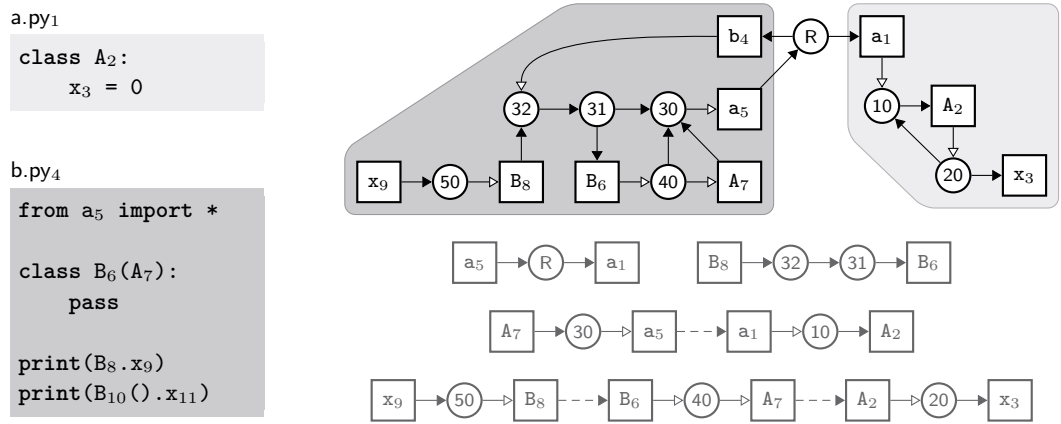
\begin{figure}[tb]
\centering
\begin{minipage}[c]{9.5em}
\begin{lstlisting}[language=python,title={\lstfilename{a.py$\sid{1}$}},backgroundcolor=\color{codea}]
class A?\sid{2}?:
    x?\sid{3}? = 0
\end{lstlisting}

\begin{lstlisting}[language=python,title={\lstfilename{b.py$\sid{4}$}},backgroundcolor=\color{codeb}]
from a?\sid{5}? import *

class B?\sid{6}?(A?\sid{7}?):
    pass

print(B?\sid{8}?.x?\sid{9}?)
print(B?\sid{10}?().x?\sid{11}?)
\end{lstlisting}
\end{minipage}
\hspace{2em}
\begin{tikzpicture}[baseline=(current bounding box.center)]

    \node (scope graph) [draw=none] {\begin{tikzpicture}
        \node (root) [int scope] {R};
        \node (a1) [old symbol, right=of root] {\OldSymbol{a}{1}};
        \node (s10) [int scope, below=of a1] {10};
        \node (A2) [old symbol, right=of s10] {\OldSymbol{A}{2}};
        \node (s20) [int scope, below=of A2] {20};
        \node (x3) [old symbol, right=of s20] {\OldSymbol{x}{3}};

        \node (b4) [old symbol, left=of root] {\OldSymbol{b}{4}};
        \node (a5) [old symbol, below=of b4] {\OldSymbol{a}{5}};
        \node (s30) [int scope, left=of a5] {30};

        \node (A7) [old symbol, below=of a5] {\OldSymbol{A}{7}};
        \node (s40) [int scope, at=(A7 -| s30)] {40};
        \node (B6) [old symbol, left=of s40] {\OldSymbol{B}{6}};
        \node (s31) [int scope, at=(s30 -| B6)] {31};

        \node (B8) [old symbol, left=of B6] {\OldSymbol{B}{8}};
        \node (s32) [int scope, at=(s31 -| B8)] {32};
        \node (s50) [int scope, left=of B8] {50};
        \node (x9) [old symbol, left=of s50] {\OldSymbol{x}{9}};

        \path [-neron]
        (root) edge (a1)
        (s10) edge (A2)
        (s20) edge (s10)
        (s20) edge (x3)

        (root) edge (b4)
        (a5) edge (root)
        (s31) edge (s30)
        (s31) edge (B6)
        (s40) edge (s30)
        (A7) edge (s30)
        (s32) edge (s31)
        (B8) edge (s32)
        (x9) edge (s50)
        ;

        \path [-neron import]
        (s30) edge (a5)
        (a1) edge (s10)
        (A2) edge (s20)
        (B6) edge (s40)
        (s40) edge (A7)
        (s50) edge (B8)
        (b4) edge [out control=+(+180:1cm), in control=+(+90:1cm)] (s32)
        ;

        \begin{scope}[on background layer]
            \path [shade a]
                 ($(a1.north west) + (-0.15cm, 0.15cm)$)
              -- ($(a1.south west |- s10.south) + (-0.15cm, -0.15cm)$)
              -- ($(s20.west |- x3.south west) + (-0.15cm, -0.15cm)$)
              -- ($(x3.south east) + (0.15cm, -0.15cm)$)
              -- ($(a1.north east -| x3.north east) + (0.15cm, 0.15cm)$)
              -- cycle;

            \path [shade b]
                 ($(b4.north east) + (0.15cm, 0.15cm)$)
              -- ($(b4.north west -| s31.west) + (-0.15cm, 0.15cm)$)
              -- ($(x9.north west) + (-0.15cm, 0.15cm)$)
              -- ($(x9.south west) + (-0.15cm, -0.15cm)$)
              -- ($(A7.south east) + (0.15cm, -0.15cm)$)
              -- cycle;
        \end{scope}
    \end{tikzpicture}};

    \node (paths 1) [draw=none, below=0.1cm of scope graph] {\begin{tikzpicture}
        \node (a5) [old symbol, neron path] {\OldSymbol{a}{5}};
        \node (root) [int scope, neron path, right=of a5] {R};
        \node (a1) [old symbol, neron path, right=of root] {\OldSymbol{a}{1}};
        \path [-neron path, neron path]
        (a5) edge (root)
        (root) edge (a1)
        ;

        \node (B8) [old symbol, neron path, right=0.75cm of a1] {\OldSymbol{B}{8}};
        \node (s32) [int scope, neron path, right=of B8] {32};
        \node (s31) [int scope, neron path, right=of s32] {31};
        \node (B6) [old symbol, neron path, right=of s31] {\OldSymbol{B}{6}};
        \path [-neron path, neron path]
        (B8) edge (s32)
        (s32) edge (s31)
        (s31) edge (B6)
        ;
    \end{tikzpicture}};

    \node (paths 2) [draw=none, below=0.1cm of paths 1] {\begin{tikzpicture}
        \node (A7) [old symbol, neron path] {\OldSymbol{A}{7}};
        \node (s30) [int scope, neron path, right=of A7] {30};
        \node (a5) [old symbol, neron path, right=of s30] {\OldSymbol{a}{5}};
        \node (a1) [old symbol, neron path, right=0.65cm of a5] {\OldSymbol{a}{1}};
        \node (s10) [int scope, neron path, right=of a1] {10};
        \node (A2) [old symbol, neron path, right=of s10] {\OldSymbol{A}{2}};
        \path [-neron path, neron path]
        (A7) edge (s30)
        (s10) edge (A2)
        ;
        \path [-neron path import, neron path]
        (s30) edge (a5)
        (a1) edge (s10)
        ;
        \path [-neron path, neron path resolved]
        (a5) edge (a1)
        ;
    \end{tikzpicture}};

    \node (paths 3) [draw=none, below=0.1cm of paths 2] {\begin{tikzpicture}
        \node (x9) [old symbol, neron path] {\OldSymbol{x}{9}};
        \node (s50) [int scope, neron path, right=of x9] {50};
        \node (B8) [old symbol, neron path, right=of s50] {\OldSymbol{B}{8}};
        \node (B6) [old symbol, neron path, right=0.65cm of B8] {\OldSymbol{B}{6}};
        \node (s40) [int scope, neron path, right=of B6] {40};
        \node (A7) [old symbol, neron path, right=of s40] {\OldSymbol{A}{7}};
        \node (A2) [old symbol, neron path, right=0.65cm of A7] {\OldSymbol{A}{2}};
        \node (s20) [int scope, neron path, right=of A2] {20};
        \node (x3) [old symbol, neron path, right=of s20] {\OldSymbol{x}{3}};
        \path [-neron path, neron path]
        (x9) edge (s50)
        (s20) edge (x3)
        ;
        \path [-neron path import, neron path]
        (s50) edge (B8)
        (B6) edge (s40)
        (s40) edge (A7)
        (A2) edge (s20)
        ;
        \path [-neron path, neron path resolved]
        (B8) edge (B6)
        (A7) edge (A2)
        ;
    \end{tikzpicture}};
\end{tikzpicture}
\caption{A sample program and its Néron scope graph}
\label{fig:neron}
\end{figure}

Figure~\ref{fig:neron} shows our example Python program, which
consists of two files. The
first defines a class \OldSymbol{A}{2} containing a single class field
\OldSymbol{x}{3}.\footnote{Following the scope graph literature, we add a unique
numeric subscript to each identifier so that we can easily distinguish different
occurrences of the same identifier in the program source. These subscripts are
not part of the actual identifiers seen by the Python interpreter.} The second
file imports all of the names from \OldSymbol{a}{5}, defines a subclass
\OldSymbol{B}{6}, and then prints its inherited field twice: by accessing
it first as a class member and then as an instance member.

We also show the scope graph that we would construct for
this program according to the judgments defined by Néron et al.\
\cite{Neron2015}. Circular nodes represent \emph{scopes}, which are “minimal
program regions that behave uniformly with respect to name
resolution.”\footnote{We have been careful to label scope nodes consistently
across all of our examples: scopes with the same number in each figure represent
the same regions of the example source program. In the grand tradition of a
\textsf{BASIC} programmer choosing line numbers, we have left “gaps” in the sequence of assigned
scope numbers so that related scopes can have numbers close to each other, while
leaving room to add additional scopes in later examples.} Rectangular nodes
represent definitions and references, with edges connecting a scope to each of
the symbols defined in that scope, and connecting each reference to the scope in
which that symbol should be resolved. Scope \ExternalScope{R} is a \emph{root
node}, representing the global namespace that all Python modules belong to,
while \OldSymbol{a}{1} and \OldSymbol{b}{4} are the definitions of the modules
defined in the two files.\footnote{Note that in Python, the name of a module
does not appear explicitly in the source code, and is instead inferred from the
name of the file defining it.} Scope \ExternalScope{20} contains the class
members of
class \texttt{A}, while the edge from \ExternalScope{20} to \OldSymbol{x}{3}
indicates that \texttt{x} is one of those class members. Scope
\ExternalScope{50} represents the names that are available via the member access
operator in the first \texttt{print} statement, while the edge from
\OldSymbol{x}{9} to \ExternalScope{50} indicates that the reference must be
resolved relative to those names.

Edges between scopes represent nesting. For instance, the top-level definitions
in a Python module are evaluated sequentially, and each name is only visible
after its definition has been evaluated. This is modeled by the edges connecting
scopes \ExternalScope{30}, \ExternalScope{31}, and \ExternalScope{32}, which
represent the names visible immediately after the \texttt{import},
\texttt{class}, and first \texttt{print} statements, respectively. Scope
\ExternalScope{40} contains the class members of class \texttt{B};
the edge between it and \ExternalScope{30} models how references within the
class body can refer to definitions in the containing lexical scope. Open-arrow
edges represent “imports,” which make the contents of a named scope available
elsewhere. For instance, the edge between \ExternalScope{40} and
\OldSymbol{A}{7} specifies that \texttt{B} inherits all of the class members of
\texttt{A}, while the edge between \OldSymbol{A}{2} and \ExternalScope{20}
specifies that scope \ExternalScope{20} contains the class members of class
\texttt{A}.

We also show the name binding path that correctly resolves the reference
\OldSymbol{x}{9} to its definition \OldSymbol{x}{3}. This requires resolving
several intermediate references via import edges: resolving \OldSymbol{B}{8} to
the definition of class \texttt{B}; \OldSymbol{A}{7} to its superclass; and
\OldSymbol{a}{5} to the module defining that superclass.

Importantly, the Néron scope graph can be divided into disjoint subgraphs for
each file, indicated by shading. Apart from the root node (which is a singleton
node shared across all files), every node belongs to some file's subgraph. Every
edge connects two nodes that belong to the same file, or connects a node to the
shared root node. At query time, we must consider the scope graph as a whole,
since name binding paths can cross from one subgraph to another. However, we can
\emph{construct} each file's subgraph at index time without having to consider
the content of any other file in the program. Néron scope graph construction is
therefore file-incremental in exactly the way that we need.

\subsection{Van~Antwerpen scope graphs}

Unfortunately, the Néron scope graph does not allow us to resolve
\OldSymbol{x}{11}, since it depends on \emph{type-dependent name resolution}. To
resolve \OldSymbol{x}{11}, we must know the return type that we get from
invoking \OldSymbol{B}{10}, so that we can resolve the reference in the return
type's scope. The return type will depend on what kind of entity
\OldSymbol{B}{10} resolves to, which we do not know \emph{a priori}. In this
particular example, it resolves to a class definition, and so the return type is
an instance of that class. But it could resolve to a function (or any other
“callable”), with an arbitrary return type defined by the function body.

\begin{figure}[tb]
\centering
\begin{tikzpicture}

    \node (root) [int scope] {R};

    \node (a1) [old symbol, right=0.75cm of root] {$\OldSymbol{a}{1}: \Rec{10}$};
    \node (s10) [int scope, below=of a1] {10};
    \node (A2) [old symbol, right=0.75cm of s10] {$\OldSymbol{A}{2}: \Class{20}$};
    \node (A2c) [old symbol, right=0.75cm of a1] {$\OldSymbol{A}{2}: \Instance{21}$};
    \node (s20) [int scope, below=of A2] {20};
    \node (x3) [old symbol, below=0.75cm of s20] {$\OldSymbol{x}{3}: \Num$};
    \node (s21) [int scope, right=0.75cm of s20] {21};

    \node (b4) [old symbol, left=0.75cm of root] {$\OldSymbol{b}{4}: \Rec{33}$};
    \node (a5) [old symbol, below=of b4.south east, anchor=north east] {\OldSymbol{a}{5}};
    \node (x9) [old symbol, below=of a5.south east, anchor=north east] {\OldSymbol{x}{9}};
    \node (x11) [old symbol, below=of x9.south east, anchor=north east] {\OldSymbol{x}{11}};
    \node (B6) [old symbol, below=of x11.south east, anchor=north east] {$\OldSymbol{B}{6}: \Class{40}$};
    \node (s40) [int scope, right=of B6] {40};
    \node (s31) [int scope, above=of B6.north west, anchor=south west] {31};
    \node (s30) [int scope, above=of s31] {30};
    \node (A7) [old symbol, above=of s30] {\OldSymbol{A}{7}};
    \node (B6c) [old symbol, below=of B6.south east, anchor=north east]
                {$\OldSymbol{B}{6}: \Instance{41}$};
    \node (s41) [int scope, at=(B6c -| s40)] {41};

    \node (s32) [int scope, left=0.75cm of s31] {32};
    \node (B8) [old symbol, above=of s32] {\OldSymbol{B}{8}};
    \node (s33) [int scope, left=0.75cm of s32] {33};
    \node (B10) [old symbol, below=of s33] {\OldSymbol{B}{10}};

    \path [-hendrik link in]
    (root) edge [":" hendrik side label] (a1)
    (root) edge [":"' hendrik side label] (b4)
    (s10) edge [":" {hendrik side label, below}] (A2)
    (s10) edge ["()" {hendrik side label, above, pos=0.8, inner sep=0.5em},
               out control=+(+15:1cm), in control=+(+180:1.5cm)] (A2c)
    (s20) edge [":" {hendrik side label, right, inner sep=0.4em}] (x3)
    (s31) edge [":" {hendrik side label, above, inner sep=0.5em},
               out control=+(0:0.5cm), in control=+(+90:0.75cm)] (B6)
    (s31) edge ["()" {hendrik side label, left},
               out control=+(-150:1cm), in control=+(+180:1.7cm)] (B6c)
    ;

    \path [-hendrik link out]
    (a1) edge (s10)
    (A2) edge (s20)
    (A2c) edge [out control=+(-35:1cm), in control=+(+70:1cm)] (s21)
    (b4) edge [out control=+(+180:2cm), in control=+(+90:2cm)] (s33)
    (B6) edge (s40)
    (B6c) edge (s41)
    ;

    \path [-hendrik]
    (s21) edge (s20)
    (x9) edge (s20)
    (x11) edge [out control=+(+0:1cm), in control=+(-150:1cm)] (s20)
    (s40) edge [out control=+(+40:1.5cm), in control=+(-120:1.5cm)]
         node [pos=0.5, hendrik label] {E} (s20)
    (s41) edge [out control=+(+0:3cm), in control=+(-90:3cm)]
         node [pos=0.5, hendrik label] {E} (s21)
    (s41) edge (s40)
    (s30) edge [out control=+(+45:1.5cm), in control=+(-135:1cm)]
         node [pos=0.6, hendrik label] {I} (s10)
    (A7) edge (s30)
    (s31) edge node [hendrik label, fill=none, right, inner sep=0.5em] {P} (s30)
    (s32) edge node [hendrik label, fill=none, above, inner sep=0.4em] {P} (s31)
    (s33) edge node [hendrik label, fill=none, above, inner sep=0.4em] {P} (s32)
    (B8) edge (s32)
    (a5) edge (root)
    (B10) edge (s33)
    ;

    \begin{scope}[on background layer]
        \path [shade a]
             ($(a1.north west) + (-0.15cm, 0.15cm)$)
          -- ($(a1.south west |- s10.north) + (-0.15cm, -0.15cm)$)
          -- ($(x3.south west) + (-0.15cm, -0.15cm)$)
          -- ($(x3.south east -| A2c.south east) + (0.15cm, -0.15cm)$)
          -- ($(A2c.north east) + (0.15cm, 0.15cm)$)
          -- cycle;

        \path [shade b]
             ($(b4.north east) + (0.15cm, 0.15cm)$)
          -- ($(b4.north west -| B8.north east) + (-0.15cm, 0.15cm)$)
          -- ($(B10.north west |- B8.south west) + (-0.15cm, 0.15cm)$)
          -- ($(B10.south west |- B6c.south west) + (-0.15cm, -0.15cm)$)
          -- ($(B6c.south east -| s41.east) + (0.15cm, -0.15cm)$)
          -- ($(B6.north east -| s40.east) + (0.15cm, 0.15cm)$)
          -- ($(x9.south east) + (0.15cm, -0.15cm)$)
          -- cycle;
    \end{scope}
\end{tikzpicture}
\caption{The sample program's van~Antwerpen scope graph}
\label{fig:hendrik}
\end{figure}
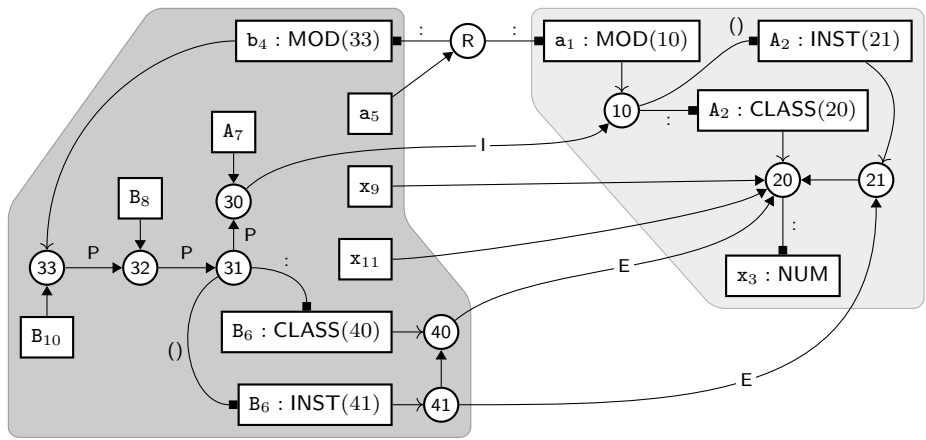

Van~Antwerpen et al.\ \cite{VanAntwerpen2016,VanAntwerpen2018} extend the scope
graph judgments to support these kinds of lookups. Figure~\ref{fig:hendrik}
shows the scope graph for our example program using these modified judgments. We
use square-cap edges to connect each scope with the definitions in that scope.
These edges are labeled to indicate the kind of definition: a \texttt{:} edge
specifies the type of a symbol defined in the scope, while a \texttt{()} edge
specifies the return type of a callable symbol defined in the scope. We need
edges with both labels for Python classes, since they can both be referred to by
name (yielding the class itself) and called (yielding an instance of the class).
Similarly, we have two associated scopes for each class: one (\ExternalScope{20}
and \ExternalScope{40}) for the class members, and one (\ExternalScope{21} and
\ExternalScope{41}) for the instance members. Each pair of scopes is connected
by an edge, modeling how class members are also available as instance members.

There are paths in the van~Antwerpen scope graph that resolve both
\OldSymbol{x}{9} and \OldSymbol{x}{11} to the definition \OldSymbol{x}{3}.
However, the paths encoding these name bindings ($\OldSymbol{x}{9} \rightarrow
\ExternalScope{20} \rightarrow \OldSymbol{x}{3}$ and $\OldSymbol{x}{11}
\rightarrow \ExternalScope{20} \rightarrow \OldSymbol{x}{3}$) are shorter than
we might expect, since they do not directly encode all of the intermediate
lookups that are needed.

The intermediate lookups do happen, but at a different time. In the Néron scope
graph, there are import edges connecting \ExternalScope{30} to \OldSymbol{a}{5}
and \OldSymbol{a}{1} to \ExternalScope{10}, which we follow lazily at
query time. In the van~Antwerpen scope graph, we perform these
intermediate lookups eagerly at graph construction time. When we
encounter the \texttt{import} statement in the source, we immediately resolve
the \OldSymbol{a}{5} reference. Doing so takes us to \OldSymbol{a}{1} and its
connected scope \ExternalScope{10}, as in the Néron scope graph. However, we
then persist this lookup result into the graph structure, as an \textsf{I} edge
directly connecting \ExternalScope{30} to \ExternalScope{10}. We perform (and
persist) similar
construction-time lookups of \texttt{B}'s superclass reference \OldSymbol{A}{7},
and of the type references \OldSymbol{B}{8} and \OldSymbol{B}{10}.

While we have gained the ability to perform type-dependent lookups, we have lost
file incrementality. The eager intermediate lookups produce edges that cross
file boundaries, violating the disjointedness property of Néron scope graphs.
This violation is not superficial; it highlights that each intermediate lookup
could depend on \emph{any other} file in the program. We cannot analyze each
file purely in isolation, since we cannot know in advance which other files we
might have to inspect. Worse, future updates to other files might invalidate the
subgraph that we create. In the pathological case, a change to one file might
require us to regenerate the scope graph for the \emph{entire program}.

\section{Stack graphs}
\label{sec:stack-graphs}

\emph{Stack graphs} support the advanced type-dependent lookups of van~Antwerpen
scope graphs. They also retain the file incrementality of Néron scope graphs, by
performing intermediate lookups lazily at query time. Our key insight is to
maintain an explicit \emph{stack} of the currently pending intermediate lookups
during the name resolution algorithm.

\begin{figure}[tb]
\centering
\begin{tikzpicture}


    \node (root1) [root node] {};
    \node (a1) [sg symbol, below=of root1] {\PopSymbol{\OldSymbol{a}{1}}};
    \node (d1) [pseudo symbol, right=of a1] {\PopSymbol{\PseudoSymbol{.}}};
    \node (c2) [pseudo symbol, right=of d1] {\PopSymbol{\PseudoSymbol{()}}};
    \node (A2) [sg symbol, below=of c2] {\PopSymbol{\OldSymbol{A}{2}}};
    \node (s10) [int scope, at=(A2 -| d1)] {10};
    \node (dc2) [pseudo symbol, right=of c2] {\PopSymbol{\PseudoSymbol{.}}};
    \node (d2) [pseudo symbol, right=of A2] {\PopSymbol{\PseudoSymbol{.}}};
    \node (s20) [int scope, right=of d2] {20};
    \node (s21) [int scope, at=(dc2 -| s20)] {21};
    \node (x3) [sg symbol, below=of s20] {\PopSymbol{\OldSymbol{x}{3}}};

    \node (b4) [sg symbol, left=0.75cm of a1] {\PopSymbol{\OldSymbol{b}{4}}};
    \node (root2) [root node, above=of b4] {};
    \node (d4) [pseudo symbol, below=of b4] {\PopSymbol{\PseudoSymbol{.}}};

    \node (s30) [int scope, left=0.6cm of root2] {30};
    \node (d5) [pseudo symbol, left=of s30] {\PushSymbol{\PseudoSymbol{.}}};
    \node (a5) [sg symbol, left=of d5] {\PushSymbol{\OldSymbol{a}{5}}};
    \node (root5) [root node, left=of a5] {};

    \node (s31) [int scope, below=0.75cm of s30] {31};
    \node (B6) [sg symbol, left=of s31] {\PopSymbol{\OldSymbol{B}{6}}};
    \node (d6) [pseudo symbol, left=of B6] {\PopSymbol{\PseudoSymbol{.}}};
    \node (c6) [pseudo symbol, below=of B6] {\PopSymbol{\PseudoSymbol{()}}};
    \node (dc6) [pseudo symbol, left=of c6] {\PopSymbol{\PseudoSymbol{.}}};
    \node (s40) [int scope, left=of d6] {40};
    \node (s41) [int scope, left=of dc6] {41};
    \node (d7) [pseudo symbol, left=of s40] {\PushSymbol{\PseudoSymbol{.}}};
    \node (dc7) [pseudo symbol, left=of s41] {\PushSymbol{\PseudoSymbol{.}}};
    \node (A7) [sg symbol, left=of d7] {\PushSymbol{\OldSymbol{A}{7}}};
    \node (c7) [pseudo symbol, at=(dc7 -| A7)] {\PushSymbol{\PseudoSymbol{()}}};

    \node (B8) [sg symbol, below=of B6.south east |- c6.south east,
                anchor=north east] {\PushSymbol{\OldSymbol{B}{8}}};
    \node (s32) [int scope, at=(s31 |- B8)] {32};
    \node (d9) [pseudo symbol, left=of B8] {\PushSymbol{\PseudoSymbol{.}}};
    \node (x9) [sg symbol, left=of d9] {\PushSymbol{\OldSymbol{x}{9}}};

    \node (B10) [sg symbol, below=of B8.south east,
                 anchor=north east] {\PushSymbol{\OldSymbol{B}{10}}};
    \node (s33) [int scope, at=(s32 |- B10)] {33};
    \node (c10) [pseudo symbol, left=of B10] {\PushSymbol{\PseudoSymbol{()}}};
    \node (d10) [pseudo symbol, left=of c10] {\PushSymbol{\PseudoSymbol{.}}};
    \node (x11) [sg symbol, left=of d10] {\PushSymbol{\OldSymbol{x}{11}}};

    \path [-sg]
    (root1) edge [sg path] (a1)
    (a1) edge [sg path] (d1)
    (d1) edge [sg path] (s10)
    (s10) edge [sg path] (A2)
    (A2) edge [sg path] (d2)
    (A2) edge (c2)
    (d2) edge [sg path] (s20)
    (s20) edge [sg path] (x3)
    (c2) edge (dc2)
    (dc2) edge (s21)
    (s21) edge (s20)

    (root2) edge (b4)
    (b4) edge (d4)
    (d4) edge [out control=+(-90:1cm), in control=+(+45:1cm)] (s33)
    (x11) edge [sg path] (d10)
    (d10) edge [sg path] (c10)
    (c10) edge [sg path] (B10)
    (B10) edge [sg path] (s33)
    (s33) edge [sg path] (s32)
    (x9) edge (d9)
    (d9) edge (B8)
    (B8) edge (s32)
    (s32) edge [sg path] (s31)
    (s31) edge [sg path] (B6)
    (B6) edge (d6)
    (d6) edge (s40)
    (s40) edge [sg path] (d7)
    (d7) edge [sg path] (A7)
    (B6) edge [sg path] (c6)
    (c6) edge [sg path] (dc6)
    (dc6) edge [sg path] (s41)
    (s41) edge [sg path] (s40)
    (s41) edge (dc7)
    (dc7) edge (c7)
    (c7) edge (A7)
    (A7) edge [sg path] [out control=+(+45:1cm), in control=+(-135:1cm)] (s30)
    (s31) edge (s30)
    (s30) edge [sg path] (d5)
    (d5) edge [sg path] (a5)
    (a5) edge [sg path] (root5)

    (root5) edge [out control=+(+45:1.25cm), in control=+(+135:1.25cm), overlay]
                 [sg path, densely dashed] (root1)
    ;

    \begin{scope}[on background layer]
        \path [shade a]
             ($(a1.north west |- a5.north west) + (-0.15cm, 0.15cm)$)
          -- ($(a1.south west) + (-0.15cm, -0.15cm)$)
          -- ($(c2.south west |- x3.south west) + (-0.15cm, -0.15cm)$)
          -- ($(x3.south east) + (0.15cm, -0.15cm)$)
          -- ($(a5.north east -| x3.north east) + (0.15cm, 0.15cm)$)
          -- cycle;

        \path [shade b]
             ($(b4.north east |- a5.north east) + (0.15cm, 0.15cm)$)
          -- ($(b4.south east |- B10.south east) + (0.15cm, -0.15cm)$)
          -- ($(A7.south west |- x11.south west) + (-0.15cm, -0.15cm)$)
          -- ($(A7.north west) + (-0.15cm, 0.15cm)$)
          -- ($(root5.north west |- a5.north west) + (-0.15cm, 0.15cm)$)
          -- cycle;
    \end{scope}
\end{tikzpicture}
\vskip 4pt
\begin{ceqn}
\begin{equation*}
    \begin{array}[t]{@{}l@{\:}c@{\:}ll@{\hskip 6pt}c@{}}
        \PushSymbol{\OldSymbol{x}{11}}
        & \leadsto & \PushSymbol{\OldSymbol{x}{11}}
          & \langle \sym{x} \rangle \\
        & \leadsto & \PushSymbol{\PseudoSymbol{.}}
          & \langle \sym{.} \sym{x} \rangle \\
        & \leadsto & \PushSymbol{\PseudoSymbol{()}}
          & \langle \sym{()} \sym{.} \sym{x} \rangle \\
        & \leadsto & \PushSymbol{\OldSymbol{B}{10}}
          & \langle \sym{B} \sym{()} \sym{.} \sym{x} \rangle \\
        & \leadsto & \ExternalScope{33}
          & \langle \sym{B} \sym{()} \sym{.} \sym{x} \rangle
          & \spathstep{1} \\
        & \leadsto & \ExternalScope{32}
          & \langle \sym{B} \sym{()} \sym{.} \sym{x} \rangle \\
        & \leadsto & \ExternalScope{31}
          & \langle \sym{B} \sym{()} \sym{.} \sym{x} \rangle \\
        & \leadsto & \PopSymbol{\OldSymbol{B}{6}}
          & \langle \sym{()} \sym{.} \sym{x} \rangle \\
        & \leadsto & \PopSymbol{\PseudoSymbol{()}}
          & \langle \sym{.} \sym{x} \rangle \\
    \end{array}
    \hskip 3em
    \begin{array}[t]{@{}l@{\:}c@{\:}ll@{\hskip 6pt}c@{}}
        & \leadsto & \PopSymbol{\PseudoSymbol{.}}
          & \langle \sym{x} \rangle \\
        & \leadsto & \ExternalScope{41}
          & \langle \sym{x} \rangle
          & \spathstep{2} \\
        & \leadsto & \ExternalScope{40}
          & \langle \sym{x} \rangle \\
        & \leadsto & \PushSymbol{\PseudoSymbol{.}}
          & \langle \sym{.} \sym{x} \rangle \\
        & \leadsto & \PushSymbol{\OldSymbol{A}{7}}
          & \langle \sym{A} \sym{.} \sym{x} \rangle \\
        & \leadsto & \ExternalScope{30}
          & \langle \sym{A} \sym{.} \sym{x} \rangle
          & \spathstep{3} \\
        & \leadsto & \PushSymbol{\PseudoSymbol{.}}
          & \langle \sym{.} \sym{A} \sym{.} \sym{x} \rangle \\
        & \leadsto & \PushSymbol{\OldSymbol{a}{5}}
          & \langle \sym{a} \sym{.} \sym{A} \sym{.} \sym{x} \rangle \\
        & \leadsto & \RootNode
          & \langle \sym{a} \sym{.} \sym{A} \sym{.} \sym{x} \rangle
          & \spathstep{4} \\
    \end{array}
    \hskip 3em
    \begin{array}[t]{@{}l@{\:}c@{\:}ll@{\hskip 6pt}c@{}}
        & \leadsto & \PopSymbol{\OldSymbol{a}{1}}
          & \langle \sym{.} \sym{A} \sym{.} \sym{x} \rangle \\
        & \leadsto & \PopSymbol{\PseudoSymbol{.}}
          & \langle \sym{A} \sym{.} \sym{x} \rangle \\
        & \leadsto & \ExternalScope{10}
          & \langle \sym{A} \sym{.} \sym{x} \rangle
          & \spathstep{5} \\
        & \leadsto & \PopSymbol{\OldSymbol{A}{2}}
          & \langle \sym{.} \sym{x} \rangle \\
        & \leadsto & \PopSymbol{\PseudoSymbol{.}}
          & \langle \sym{x} \rangle \\
        & \leadsto & \ExternalScope{20}
          & \langle \sym{x} \rangle
          & \spathstep{6} \\
        & \leadsto & \PopSymbol{\OldSymbol{x}{3}}
          & \phantom{\langle} \cyn \\
    \end{array}
\end{equation*}
\end{ceqn}
\caption{The sample program's stack graph}
\label{fig:stack-graph}
\end{figure}
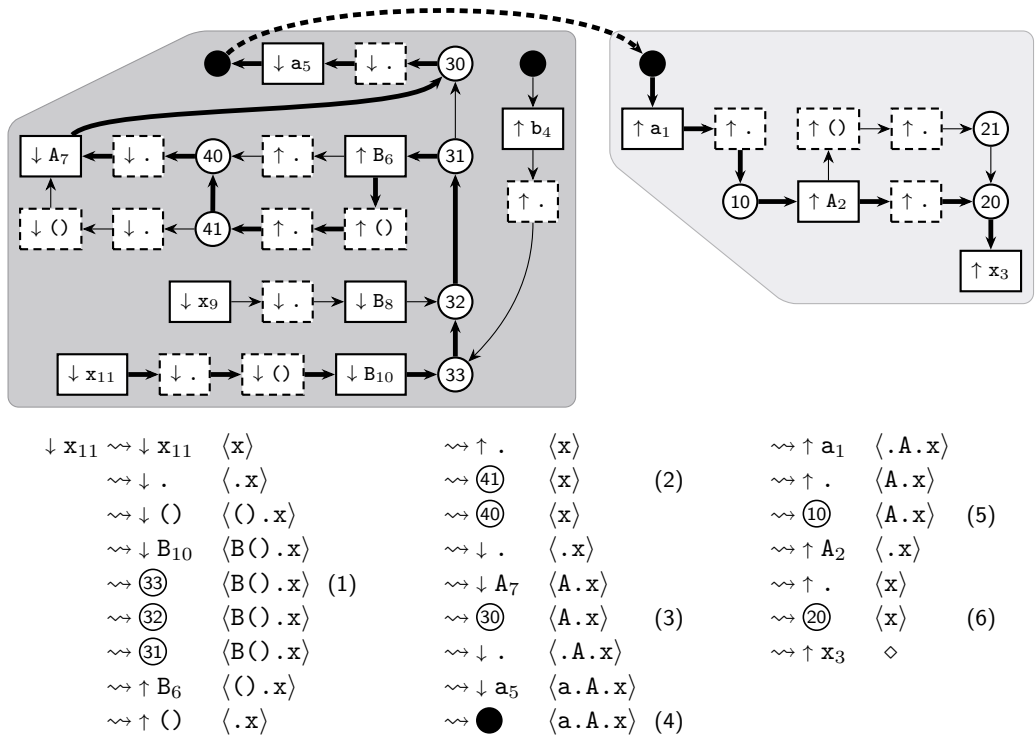

Figure~\ref{fig:stack-graph} shows the stack graph for our example program.
There is no longer a shared singleton root node; the graph can contain multiple
root nodes, which are indicated by filled circles. Definition and reference
nodes
have a solid border. Nodes with a dashed border are \emph{push} and \emph{pop}
nodes; we will see below how they enable type-dependent lookups. We add arrows
to clearly distinguish definition and pop nodes ($\uparrow$) from reference and
push nodes ($\downarrow$).

We also highlight a name binding path that resolves \OldSymbol{x}{11} to
\OldSymbol{x}{3}, and trace the discovery of that path.
We start with an “empty” path from
\OldSymbol{x}{11} to itself. With each step, we append an edge to the path, and
show the path's current frontier and the stack of currently pending lookups.
This stack contains both program identifiers (e.g. \sym{x}) indicating names
that we need to resolve, and operators (e.g. \sym{.}) indicating how the
resolved definitions will be used. Whenever we encounter a reference or push
node, we prepend the node's symbol to the stack. Whenever we encounter a
definition or pop node, we verify that the stack starts with the node's symbol,
and remove it from the stack. The final name binding path ends at a definition
node and has an empty stack, indicating that there are no more pending lookups,
and the name binding path is complete.

Several steps of interest are highlighted. At step \pathstep{1} we have seeded
the stack with a representation of the full expression being resolved. The top
of the stack specifies that the first intermediate lookup that we must perform is to
resolve \OldSymbol{B}{10}. The path's frontier indicates that this initial
lookup is to be performed in the context of scope \ExternalScope{33}, which
represents the names that are visible at the end of the module.

At step \pathstep{2} we have resolved \OldSymbol{B}{10}, and have just popped
off the \sym{()} and \sym{.} operator symbols to determine what occurs when we
invoke its definition and perform member access on the result. The stack still
contains \sym{x}, which will be resolved next. The path's frontier is
\ExternalScope{41}, indicating that \OldSymbol{x}{11} will be resolved as an
instance member of \sym{B}. The very next edge takes us to \ExternalScope{40},
which allows us to attempt to resolve \OldSymbol{x}{11} as a class member,
remembering that the instance members of a class include all of its class
members.

At step \pathstep{3} we have followed the edges modeling the class inheritance.
Doing so pushes additional symbols onto the stack, indicating that we might find
\sym{x} as an inherited class member of \sym{A}. Our next step is to resolve
\OldSymbol{A}{7} in the context of the path's frontier—scope \ExternalScope{30},
which represents the names that are visible immediately before \sym{B}'s class
definition.

At step \pathstep{4} we have followed the edges representing the \texttt{import}
statement. The stack now describes how \emph{any} symbol that we are currently
looking for might be imported from module \sym{a}. The path's frontier is a
root node. Root nodes are the only way that a name binding path can cross from
one file to another: when we encounter a root node, we can add a \emph{virtual
edge} (the dashed edge in Figure~\ref{fig:stack-graph}) to any other root node,
in any file.

At step \pathstep{5} we have resolved \OldSymbol{a}{5} to scope
\ExternalScope{10}, which represents the definition of the imported module. At
step \pathstep{6}, we have resolved \OldSymbol{A}{7} to scope
\ExternalScope{20}, which contains the class members of class \sym{A}. From
there, all that remains is to resolve \OldSymbol{x}{11} to its definition,
\OldSymbol{x}{3}.

\begin{figure}[tb]
\begin{ceqn}
\begin{equation*}
    \begin{array}[t]{@{}rc@{}}
        \textit{stack graph} & G \\
        \textit{source program} & P \\
        \textit{symbol} & x \\
        \textit{node identifier} & i \\
        \textit{source file} & \nodefile{G}{i} \
    \end{array}
    \hskip 3em
    \begin{gathered}[t]
    \begin{array}[t]{@{}rr@{\:\:}c@{\:\:}l@{}}
        \textit{node} & \graphnode{G}{i} & := &
        \RootNode \alt
        \InternalScope \alt
        \PushSymbol{x} \alt
        \PopSymbol{x} \\
        \textit{edge} & \graphedges{G} & \ni & \edge{i}{i'}{p} \\
        \textit{symbol stack} & \cys & := &
          \cyn \alt \cye \cdot \cys \\
        \textit{path} & p & := &
          \stackgraphpath{i}{i'}{\cys}{\ccs} \\
    \end{array} \\[0.5em]
    \edge{i}{i'}{p} \in \graphedges{G}
    \Rightarrow
    \nodefile{G}{i} = \nodefile{G}{i'}
    \end{gathered}
\end{equation*}
\end{ceqn}
\caption{Stack graph structure and paths}
\label{fig:stack-graph-math}
\end{figure}

The formal definition of a stack graph is shown in
Figure~\ref{fig:stack-graph-math}. A stack graph $\stackgraph{G}$ is a
representation of a \emph{source program} $P$, which consists of a set of
\emph{source files}. Each source file can be parsed into a set of \emph{syntax
nodes}. A subset of those syntax nodes represent \emph{definitions}, and a
different (not necessarily disjoint) subset of syntax nodes represent
\emph{references}. A \emph{symbol} $x$ is an identifier from the source
language, representing (part of) the name of an entity in the program, or an
operator symbol like \sym{.} or \sym{()}.

Each node in a stack graph has a unique \emph{node identifier}, and must be one
of the following: a \emph{root} node $\RootNode$, a \emph{scope} node
$\InternalScope$, a \emph{push symbol} node $\PushSymbol{x}$, or a \emph{pop
symbol} node $\PopSymbol{x}$. $\graphnode{G}{i}$ denotes the node in
$\stackgraph{G}$ with node identifier $i$. Each node \emph{belongs to} exactly
one source file, denoted $\nodefile{G}{i}$. Some nodes belong to a specific
syntax node within that file. A pop symbol node is a \emph{definition node} if
it belongs to a syntax node that is a definition. A push symbol node is a
\emph{reference} if it belongs to a syntax node that is a reference.
$\graphedges{G}$ denotes the set of edges in stack graph $\stackgraph{G}$.
Each edge $e$ in a stack graph is directed, connecting a \emph{source node} $i$
to a \emph{sink node} $i'$.
Edges can only connect nodes that belong to the same file.

\begin{figure}[tb]
\begin{equation*}
\begin{array}{@{}c@{\qquad}l}
\begin{array}{@{}c}
    \inferrule[LiftPush]
      {\graphnode{G}{i} = \PushSymbol{x}}
      {\validpath{G}{\stackgraphpath{i}{i}{x}{\ccn}}}
      \\[1.5\mathgap]
    \inferrule[LiftNoop]
      {\graphnode{G}{i} \in \{\RootNode, \InternalScope\}}
      {\validpath{G}{\stackgraphpath{i}{i}{\cyn}{\ccn}}}
\end{array}
      &
\begin{array}{@{}l@{\;\;}c}
    \rulename{Noop} &
    \inferrule
      {
          \validpath{G}{\stackgraphpath{i_0}{i_1}{\cys}{\ccs}} \\
          \edge{i_1}{i_2}{p} \in \graphedges{G} \\
          \graphnode{G}{i_2} \in \{\RootNode, \InternalScope\}
      }
      {\validpath{G}{\stackgraphpath{i_0}{i_2}{\cys}{\ccs}}}
    \\[\mathgap]
    \rulename{Push} &
    \inferrule
      {
          \validpath{G}{\stackgraphpath{i_0}{i_1}{\cys}{\ccs}} \\
          \edge{i_1}{i_2}{p} \in \graphedges{G} \\
          \graphnode{G}{i_2} = \PushSymbol{x}
      }
      {\validpath{G}{\stackgraphpath{i_0}{i_2}{x \cdot \cys}{\ccs}}}
    \\[\mathgap]
    \rulename{Pop} &
    \inferrule
      {
          \validpath{G}{\stackgraphpath{i_0}{i_1}{x \cdot \cys}{\ccs}} \\
          \edge{i_1}{i_2}{p} \in \graphedges{G} \\
          \graphnode{G}{i_2} = \PopSymbol{x}
      }
      {\validpath{G}{\stackgraphpath{i_0}{i_2}{\cys}{\ccs}}}
    \\[\mathgap]
    \rulename{Root} &
    \inferrule
      {
          \validpath{G}{\stackgraphpath{i_0}{i_1}{\cys}{\ccs}} \\
          \graphnode{G}{i_1} = \RootNode \\
          \graphnode{G}{i_2} = \RootNode \\
          i_1 \neq i_2
      }
      {\validpath{G}{\stackgraphpath{i_0}{i_2}{\cys}{\ccs}}}
\end{array}
\end{array}
\end{equation*}
\caption{Constructing paths by lifting nodes and appending edges}
\label{fig:constructing-paths}
\end{figure}

The judgment $\validpath{G}{p}$ states that $p$ is a valid path in stack graph
$G$. A path consists of a \emph{start node} $i$, an \emph{end node} $i'$, and
a \emph{symbol stack} $\cys$. A path is \emph{complete} if its start node is a
reference node, its end node is a definition node, and its symbol stack is
empty. Every name binding in a source program $P$ is represented by a complete
path in the corresponding stack graph $G$.

Paths are constructed according to the rules in
Figure~\ref{fig:constructing-paths}. An \emph{empty} path is one with no edges.
It is created by \emph{lifting} a stack graph node. Push symbol nodes “seed” the
path's symbol stack. Pop symbol nodes cannot be lifted into paths. All other
nodes result in an empty path with an empty symbol stack. We can extend a path
by \emph{appending} any edge whose start node is the same as the path's end
node. The path's symbol stack might change, depending on the edge's end node.
Root nodes and scope nodes are “noops,” which leave the symbol stack unchanged.
Push symbol nodes prepend an element onto the symbol stack. Pop symbol nodes act
as a “guard,” requiring that the top of the symbol stack match the node's
symbol. This symbol is removed from the symbol stack. (This explains why pop
symbol nodes cannot be lifted into empty paths, since there is no symbol stack
yet to satisfy this constraint.) If a path ends at a root node, we can
immediately extend it to any other root node in the stack graph via a
\emph{virtual edge}. (This is the only way that paths cross from one file to
another.)

Given these path construction judgments, we can implement a \emph{jump to
definition} algorithm: given a source program, and a reference in the program,
we want to find all of the definitions that the reference resolves
to.\footnote{When compiling or executing a program, the source language's
semantics will typically require that each reference resolve to \emph{exactly
one} definition. For an exploratory feature like code navigation, we loosen this
restriction to allow ambiguous and missing bindings. This lets us present useful
information even in the presence of incorrect programs or an incomplete model of
the language's semantics.} First, we construct the stack graph for the program.
We then perform a breadth-first search of the graph, maintaining a set of
\emph{pending} paths, starting with the empty path we get from lifting the
reference node. We use a fixed-point loop to find new pending paths by appending
(possibly virtual) edges to existing ones. When appending edges, we must satisfy
the constraints of the path construction judgments in
Figure~\ref{fig:constructing-paths}, and be careful to detect cycles. Whenever
we encounter a path that is complete, its end node identifies one of the
definitions that the reference resolves to.

This algorithm can be divided such that we construct the stack graph at index
time, and save a representation of the graph to persistent storage. At query
time, we load in the stack graph and perform the path-finding search. This
allows us to amortize the cost of constructing the stack graph across multiple
queries. Moveover, stack graph construction is file-incremental, since each node
and (non-virtual) edge belongs to exactly one file. By structuring the
persistent representation of the stack graph so that each file's subgraph can be
identified and stored independently, we can track which file versions we have
already created subgraphs for. When we receive a new commit for a repository at
index time, we only have to parse and generate new subgraphs for the file
versions not seen in previously indexed commits.

\section{Optimizing queries using partial paths}
\label{sec:partial-paths}

The process described in §\ref{sec:stack-graphs} splits work between index time
and query time, and ensures that the work we do at index time is
file-incremental, with all \emph{non}-file-incremental work happening at query
time. Unfortunately, this process performs \emph{too much} work at query time.
The path-finding search is not cheap, and as users perform many queries over
time, we will duplicate the work of discovering the overlapping parts of the
resulting name binding paths.

We would like to shift some of this work back to index time, while ensuring that
all index-time work remains file-incremental. To do this, we calculate
\emph{partial paths} for each file, which precompute portions of the
path-finding search. Because stack graphs have limited places where a path can
cross from one file into another, we can calculate all of the possible partial
paths that remain within a single file, or which reach an “import/export” point.

\begin{figure}[tb]
\begin{ceqn}
\begin{equation*}
    \begin{gathered}[c]
    \begin{array}[t]{@{}rc@{\:\:}c@{\:\:}l@{}}
        \textit{symbol stack variable} & \pyv \\
        \textit{partial symbol stack} & \pys & := &
        \pyp \alt
        \pyp \cdot \pyv \\
        \textit{partial path} & \tilde{p} & := &
        \stackgraphpartialpath{i}{\pys}{\pcs}{i'}{\pys'}{\pcs'} \\
    \end{array} \\[0.5em]
    \lift{\stackgraphpath{i}{i'}{\cys}{\ccs}} \defeq
    \stackgraphpartialpath
    {i}{\pyn}{\pcn}
    {i'}{\cys}{\ccs} \\
    \end{gathered}
    \qquad
    \begin{array}[c]{@{}r@{\;}c@{\,}c@{\,}c@{\;}l@{}}
    \stackgraphpartialpath
        {&\PushSymbol{\OldSymbol{x}{11}}&}{\pyn}{\pcn}
        {&\ExternalScope{33}&}{\langle \sym{B} \sym{()} \sym{.} \sym{x} \rangle}{\pcn} \\
    \stackgraphpartialpath
        {&\ExternalScope{33}&}{\langle \sym{B} \sym{()} \sym{.} \rangle \cdot \pyv}{\pcn}
        {&\ExternalScope{41}&}{\pyv}{\pcn} \\
    \stackgraphpartialpath
        {&\ExternalScope{41}&}{\pyv}{\pcn}
        {&\ExternalScope{30}&}{\langle \sym{A} \sym{.} \rangle \cdot \pyv}{\pcn} \\
    \stackgraphpartialpath
        {&\ExternalScope{30}&}{\pyv}{\pcn}
        {&\RootNode&}{\langle \sym{a} \sym{.} \rangle \cdot \pyv}{\pcn} \\
    \stackgraphpartialpath
        {&\RootNode&}{\langle \sym{a} \sym{.} \rangle \cdot \pyv}{\pcn}
        {&\ExternalScope{10}&}{\pyv}{\pcn} \\
    \stackgraphpartialpath
        {&\ExternalScope{10}&}{\langle \sym{A} \sym{.} \rangle \cdot \pyv}{\pcn}
        {&\ExternalScope{20}&}{\pyv}{\pcn} \\
    \stackgraphpartialpath
        {&\ExternalScope{20}&}{\langle \sym{x} \rangle \cdot \pyv}{\pcn}
        {&\PopSymbol{\OldSymbol{x}{3}}&}{\pyv}{\pcn} \\
    \end{array}
\end{equation*}
\end{ceqn}
\caption{Partial paths}
\label{fig:partial-paths}
\end{figure}

Partial paths are defined in Figure~\ref{fig:partial-paths}. A partial path
consists of a \emph{start node} $i$, an \emph{end node} $i'$, and a
\emph{precondition} $\pys$ and \emph{postcondition} $\pys'$, which are each
partial symbol stacks. A \emph{partial symbol stack} is a symbol stack with an
optional \emph{symbol stack variable}. (Note that a symbol stack variable can
only appear at the \emph{end} of a partial symbol stack.) Every path has an
equivalent partial path whose precondition is empty, and whose postcondition is
the path's symbol stack.

Partial paths are constructed using \emph{lift} and \emph{append} judgments
similar to those for paths. Partial paths can be \emph{concatenated} by unifying
the left-hand side's postcondition with the right-hand side's precondition, and
substituting any resulting symbol stack variable assignments into the left-hand
side's precondition and right-hand side's postcondition.

Figure~\ref{fig:partial-paths} also shows several example partial paths. Each
partial path precomputes a portion of the name binding path that we traced
through in Figure~\ref{fig:stack-graph}, starting and/or ending at one of the
highlighted steps. These partial paths can be concatenated together, yielding
the complete name binding path from \OldSymbol{x}{11} to \OldSymbol{x}{3}.

Partial paths give us a better balance of work between index time and query
time. At index time, we parse each previously unseen source file and produce a
stack subgraph for it, as before. We then find all partial paths within the file
between certain \emph{endpoint} nodes (root nodes, definition and reference
nodes, and certain important scope nodes). Instead of saving the subgraph
structure to persistent storage, we save this list of partial paths. At query
time, our algorithm has the same overall structure as before. Instead of
tracking pending paths and appending compatible edges to them, we track pending
partial paths and concatenate them with compatible \emph{partial path
extensions}. This process is guaranteed to find all name binding paths that
satisfy the path construction judgments from
§\ref{sec:stack-graphs}.\footnote{Due to space limitations, we do not provide
full definitions of partial path construction or concatenation or a proof of
this guarantee. These will appear in a future paper.} Like our previous
algorithm, we amortize the cost of stack graph construction across multiple
queries by performing it once at index time. By precalculating partial paths at
index time, the new algorithm also amortizes large parts of the path-finding
search.

\section{Discussion}
\label{sec:discussion}

In this section we describe how stack graphs relate to other work in this area.
The most obvious comparison is with scope graphs. In §\ref{sec:stack-graphs} we
described how stack graphs were inspired by scope graphs, and were
developed in an attempt to support type-dependent lookups while retaining file
incrementality.

Ours is not the only attempt to add incrementality to an existing program
analysis. Other approaches typically focus on what we term \emph{delta
incrementality} \cite{Kloppenburg2009,LeDilavrec2022}. Given a full result set,
one determines which files each result depends on. When a file is then changed,
only a subset of results are invalidated and recomputed.

Zwaan \cite{Zwaan2022} shows how to add delta incrementality to van~Antwerpen
scope graphs. In our running example, this approach can detect if an edit
potentially causes \OldSymbol{x}{11} to resolve to some other definition. If so,
the entire file is reanalyzed. The updated scope graph still has edges that
cross file boundaries, whose sink nodes might have changed due to edits in other
files. This means that we must store separate copies of each file's scope
subgraph for each commit that the file version appears in. The resulting storage
costs would be prohibitive. In contrast, because stack graph content is
file-incremental, each file's subgraph can be stored once, in isolation, and
reused however many times that file version appears in the project's history.
Moreover, we can detect skippable precalculation work early, using only the blob
identifiers provided by \git.

The problem of language support is not unique to large software forges like
GitHub. Local editors have coalesced around the Language Server Protocol (LSP)
standard \cite{lsp}, which provides an abstraction barrier between
language-specific and editor-specific tooling. Instead of requiring $L \times E$
integrations (one for every combination of language and editor), there are $L$
“server” implementations and $E$ “client” implementations. This greatly reduces
the total amount of development work needed to support code navigation
“everywhere.”

An LSP server typically runs alongside the local editor as an interactive
“sidecar” process, answering query requests triggered by user interactions in
the editor. This design makes LSP servers unsuitable for large software forges,
since it would require maintaining a large enough fleet of LSP servers to handle
the maximum expected number of simultaneous user queries. Moreover, we would
have to maintain separate fleets of LSP servers for each supported language. The
operational burden of maintaining these fleets counteracts the development time
saved by reusing existing LSP implementations.

In response to these difficulties, the LSP community developed the Language
Server Index Format (LSIF) \cite{lsif}. This allows LSP servers to run in
“batch” mode, producing a list of all name binding resolutions, which can be
saved into a simple database table for easy and fast querying at any point in
the future. While the LSIF specification is \emph{capable} of storing
incremental results from analyzing individual files, no extant LSP
\emph{implementations} support that mode of operation. Instead, they are run
against an entire project snapshot, typically in the same continuous integration
(CI) pipeline used to build and test the project. When a new commit arrives, the
entire project needs to be reanalyzed, even if only a small number of files have
changed.\footnote{LSP servers typically piggy-back on
existing compilers and linters, which historically have not worried about
incremental operation. Updating the LSP server to be incremental 
would require retrofitting the existing compiler, which is a more substantial
undertaking than writing an incremental LSP server from scratch
\cite{Smits2020}.}

LSP and LSIF are the standardization of many previous language-specific
frameworks for generating code navigation data at build time. All build-time
analyses suffer the same drawbacks. They require the package owner to explicitly
specify how to analyze their project,\footnote{Modern languages tend to provide
official package managers and build tools, and often admit good heuristics for
how to build a typical project. But this is not generally true for all
languages.} and by running the analysis as part of CI, require the project owner
to pay for the compute resources used.

Stack graphs, on the other hand, can be produced via a purely syntactic process,
since all name binding logic is encoded in the resulting graph structure. To
support this, we have created a new declarative \emph{graph construction
language} called \tsg\ \cite{TreeSitterGraph}. Building on the \ts\ parsing
framework \cite{TreeSitter}, the language expert defines patterns that match
against the language's grammar, and which “gadgets” of stack graph nodes and
edges should be created for each instance of those patterns in the concrete
syntax tree of a source file. These patterns are defined \emph{once} for each
language. As a result, stack graph construction requires no configuration by the
project owner, and does not need to invoke a project-specific, typically slow,
build process.

\section{Conclusion}
\label{sec:conclusion}

In this paper we have described \emph{stack graphs}, which build upon Visser et
al.'s scope graphs framework. Stack graphs can be used to perform type-dependent
name resolution, and are file-incremental, which is essential to operate
efficiently and cost-effectively at GitHub's scale. Stack graphs have been
running in production since November 2021, analyzing every commit to every
public and private Python repository hosted on GitHub. The core name resolution
algorithm is language-agnostic, and is implemented (and tested, vetted, and
deployed to production) \emph{once}. The \ts\ \cite{TreeSitter} and \tsg\
\cite{TreeSitterGraph} libraries, the stack graph algorithms \cite{StackGraphs,TreeSitterStackGraphs},
and our initial language-specific stack graph construction rulesets
\cite{StackGraphsTypescript} are all open source. This allows other tools to
incorporate stack graph code navigation features, and allows external language
communities to self-serve support for their language.

\bibliography{stack-graphs}

\end{document}